\documentclass[preprint,12pt]{elsarticle}

\usepackage{graphicx,hyperref,color,soul}
\usepackage{amssymb,amsmath}

\journal{Physica A}

\begin{document}

\begin{frontmatter}


\title{Evidence of local equilibrium in a non-turbulent Rayleigh-B{\'e}nard convection at steady-state}



\author[a]{Atanu Chatterjee}
\address[a]{Department of Physics of Complex Systems, Weizmann Institute of Science, Rehovot 7610001, Israel}
\author[b]{Takahiko Ban}
\address[b]{Division of Chemical Engineering, Department of Materials Engineering Science, Graduate School of Engineering Science, Osaka University, Machikaneyamacho 1-3, Toyonaka City, Osaka 560-8531, Japan}
\author[c]{Germano Iannacchione}
\address[c]{Department of Physics, Worcester Polytechnic Institute, 100 Institute Road, Worcester, MA 01605}

\begin{abstract}
An approach that extends equilibrium thermodynamics principles to out-of-equilibrium systems is based on the local equilibrium hypothesis. However, the validity of the \textit{a priori} assumption of local equilibrium has been questioned due to the lack of sufficient experimental evidence. In this paper, we present experimental results obtained from a pure thermodynamic study of the non-turbulent Rayleigh-B{\'e}nard convection at steady-state to verify the validity of the local equilibrium hypothesis. A non-turbulent Rayleigh-B{\'e}nard convection at steady-state is an excellent `model thermodynamic system' in which local measurements do not convey the complete picture about the spatial heterogeneity present in the macroscopic thermodynamic landscape. Indeed, the onset of convection leads to the emergence of spatially stable hot and cold domains. Our results indicate that these domains while break spatial symmetry macroscopically, preserves it locally that exhibit room temperature equilibrium-like statistics. Furthermore, the role of the emergent heat flux is investigated and a linear relationship is observed between the heat flux and the external driving force following the onset of thermal convection. Finally, theoretical and conceptual implications of these results are discussed which opens up new avenues in the study non-equilibrium steady-states, especially in complex, soft, and active-matter systems.
\end{abstract}


\end{frontmatter}


\section{Introduction}
Isolated systems when left to themselves approach a state of thermodynamic equilibrium spontaneously. This state of equilibrium can be uniquely characterized by a set of thermodynamic variables also known as state variables. A driven macroscopic system that can self-organize itself tends to resist this state of universal attractor by locking itself in meta-stable states of dynamical equilibrium. These meta-stable states give rise to complex structures which adapt and self-organize in response to the effects of external perturbations in the form of thermodynamic forces, flows, and currents. Some examples of spatio-temporal pattern formation through self-organization include phase-transition in magnetic systems, critical phenomena in dynamical systems, fluid-phase instabilities in thermo-fluid systems, self-assembly in molecular systems, oscillatory reactions in chemical systems, and collective behavior in biological, active matter, and social systems~\cite{cross1993pattern,jaeger2010far,ashby1991principles,national2007condensed,chatterjee2016thermodynamics,georgiev2016road,chatterjee2017aging,georgiev2017exponential}.

Although far-from-equilibrium systems are ubiquitous in nature, a serious thermodynamic treatment of their behavior remains largely an uncharted territory~\cite{jaeger2010far,van2020nonequilibrium,martyushev2006maximum, chatterjee2016thermodynamics,lucia2008probability,lucia2012maximum,cimmelli2014entropy,li2019entropic,dong2012general}. One of the several ways to extend equilibrium thermodynamics to out-of-equilibrium systems is through the assumption of local equilibrium in classical irreversible systems~\cite{glavatskiy2015local,vilar2001thermodynamics}. An approach based on the local equilibrium hypothesis formulates a macroscopic system as collection of `cells' (domains) in which rules of classical equilibrium thermodynamics are fulfilled to good approximation. This particular viewpoint dates back several decades when Milne, from an astrophysical perspective, defined local thermodynamic equilibrium in a local `cell'. He proposed the condition that the `cell' will continue to be at local thermodynamic equilibrium as long as it macroscopically absorbs and spontaneously emits radiation as if it were in radiative equilibrium in a cavity at the temperature of the matter of the `cell'~\cite{milne1928effect}. If these `cells' are well-defined, then they allow for transport of matter and energy in between them. This however has to follow under the strict constraint that the flows and currents between the `cells' do not disturb the respective individual local thermodynamic equilibria with respect to the intensive variables. Therefore, one can think of two `relaxation times' that are separated by orders of magnitude: the longer relaxation time responsible for the macroscopic evolution of the system ($\tau_M$) and the shorter relaxation time ($\tau_m$) responsible for local equilibration of a single cell. If these two relaxation times are not well separated, then the classical non-equilibrium thermodynamical concept of local thermodynamic equilibrium loses its meaning~\cite{glansdorff1971thermodynamic,vilar2001thermodynamics,gyarmati1970non,ai2010non,jou1999extended,de1962non,bodenschatz1992experiments}. The ratio between these two time scales is called the Deborah number, $De = \tau_m/\tau_M$. For $De<<1$, the local equilibrium hypothesis is fully justified because the relevant variables evolve on a large timescale and do not practically change over the time, but the hypothesis is not appropriate to describe situations characterized by $De>>1$~\cite{lebon2008understanding}.

While the validity of local equilibrium provides a useful framework to extend our understanding of the thermodynamics of classical irreversible systems, the assumption of local equilibrium is usually taken for granted. As has been previously noted by Ben-Naim, that the assumption of ``local equilibrium" is ill-founded, and that it is not clear whether it is possible to define an entropy-density function in order to obtain the ``entropy production" of the entire system~\cite{ben2018validity,ben2020entropy}. We acknowledge that these are open problems in non-equilibrium thermodynamics that are of great fundamental importance. Therefore, it would be too ambitious of a task on our part to attempt to answer them in this paper. Rather, we use this paper as an avenue to present experimental evidence in support of the local-equilibrium assumption in a prototypical driven system, the non-turbulent Rayleigh-B{\'e}nard convection at steady-state. This work builds on our previous studies where we have performed extensive thermal analysis of these convective cells, and have shown that the temperature manifold bifurcates into regions of local sources and sinks as macroscopic order emerges~\cite{chatterjee2019coexisting,chatterjee2019many,chatterjee2019overview,yadati2019spatio,yadati2020experimental,chatterjee2020time}. In this paper, we first show that the temperature distribution profiles of these localized domains that coexist together at a non-equilibrium steady-state exhibit room temperature equilibrium-like statistics. Next, we investigate the nature of the emergent heat flux as a function of external driving force. Beyond a critical temperature it is observed that the emergent heat flux scales linearly with the mean top temperature and the external driving force. The linearity in the force-flux relationship is important in the context of non-equilibrium thermodynamics as it gives quantitative support to the local equilibrium hypothesis as the Rayleigh-B{\'e}nard system undergoes an order-disorder like phase-transition (or can be more aptly described as, order-disorder phase-separation), while also allowing us to develop a meaningful definition of one of the key thermodynamic variables - temperature - in out-of-equilibrium scenarios.

We start with a discussion of the Rayleigh-B{\'e}nard convection and the Bousinessq approximation~\cite{koschmieder1993benard}. Next we present our experimental methodology in brief, and discuss the spatio-temporal analysis of the thermal statistics. Following which we discuss the emergent force-flux relationship as the Rayleigh-B{\'e}nard system approaches a steady-state. We argue that the thermal statistics thus obtained from the infrared thermometry of the top layer of the fluid film along with the linearity in the force-flux relationship provide a quantitative evidence for the local equilibrium hypothesis. Finally, we discuss the implications of this result in broad perspectives both from a theoretical and a conceptual point of view. 

\section{The Rayleigh-B{\'e}nard Convection}
The Rayleigh-B{\'e}nard convection holds a place of special interest in the scientific community~\cite{cross1993pattern,behringer1985rayleigh,bodenschatz2000recent}. It is one of the oldest and most widely used canonical examples to study pattern formation, emergence, and self-organization~\cite{cross1993pattern,jaeger2010far,glansdorff1971thermodynamic,koschmieder1993benard,heylighen2001science}. Although Rayleigh–Benard convection had been known since the early twentieth century, extensive work by Prigogine and colleagues established a critical link between self-organization and entropy production. They claimed that the creation of ordered structures in open systems is accompanied by increased ‘dissipation,' thus coining the term `dissipative structure'~\cite{prigogine1977time,nicolis1977self,meysman2010ecosystem}.

\begin{figure}[t]
    \centering
    \includegraphics{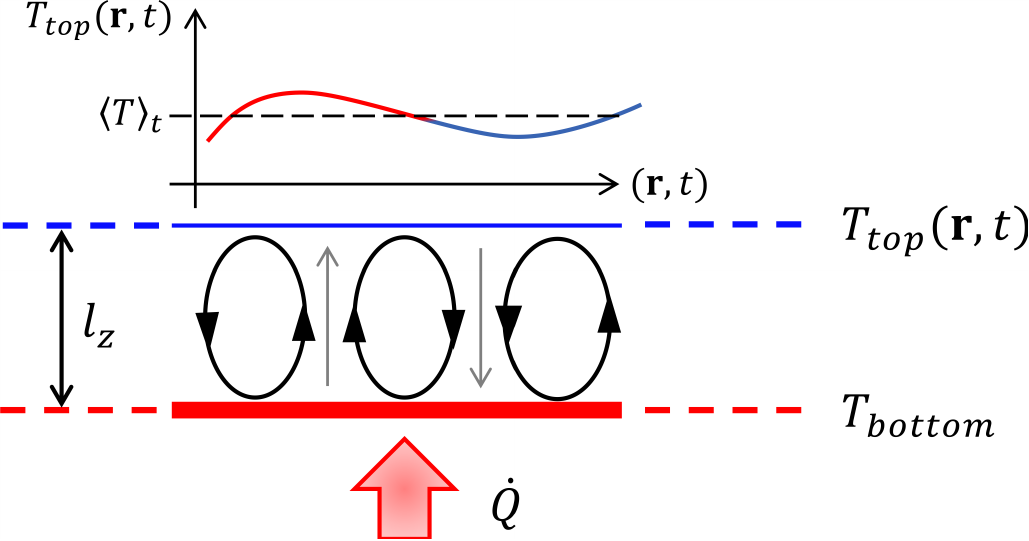}
    \caption{An illustration of the Rayleigh-B{\'e}nard system and the top layer temperature profile at steady-state under asymmetrical semi-rigid boundary condition. A thin film of viscous liquid is heated from the bottom ($\dot{Q}$). For $Ra > Ra_c$ the thermal gradient between the base and the top gives rise to convection patterns. At steady-state real-time thermal imaging of the top layer is performed to extract the spatial distribution of $T_{top}$. The line cut denotes the `thermal landscape' of the top layer due to the presence of alternating hot and cold plumes, and the dotted line represents temporally averaged top film temperature, $\langle T\rangle_t$ at steady-state.}
    \label{fig1}
\end{figure}

When a thin film of liquid is heated, the competing forces between viscosity and buoyancy give rise to convective instabilities. This convective instability creates a spatio-temporal non-uniform thermal distribution on the surface of the fluid film as shown in Figure~\ref{fig1}. The advantage of this system lies in its simplicity, wherein a dimensionless quantity, the Rayleigh number $(Ra)$, determines the onset of convective cell patterns,
\begin{equation}
    Ra = \frac{g\beta l_z^3}{\nu\alpha}(T_{bottom} - T_{top})
    \label{eqn1}
\end{equation}
Here, $l_z$ denotes fluid film thickness, $\nu$ kinematic viscosity, $\alpha$ thermal diffusivity, $\beta$ thermal expansion coefficient, and $g$ acceleration due to gravity. The critical Rayleigh number ($Ra_c$) of $1708$ marks the onset of convection for a no-slip boundary condition was obtained by Jeffreys in 1929~\cite{cross1993pattern,koschmieder1993benard,rayleigh1916lix}. Under the approximations of an ideal incompressible fluid that is thermally driven one can write the following set of equations also known as the Boussinesq approximations,
\begin{equation}
\begin{gathered}
    \frac{\partial\rho}{\partial t} + \nabla\cdot(\rho\overrightarrow{u}) = 0\\
    \frac{\partial\overrightarrow{u}}{\partial t} + (\overrightarrow{u}\cdot\nabla)\overrightarrow{u} = -\frac{1}{\rho}\nabla p + \nu\nabla^2\overrightarrow{u} - \overrightarrow{g}\beta\Delta T\\ 
    \frac{\partial T}{\partial t} + \overrightarrow{u}\cdot\nabla T = \alpha\nabla^2T + \frac{\dot{q}}{\rho c}
\end{gathered}
\label{eqn2}
\end{equation}
For a packet of fluid with local convective velocity, $\overrightarrow{u}$ incompressibility implies, $\nabla\cdot\overrightarrow{u} = 0$; the density, $\rho$ is assumed to vary linearly with temperature, $\rho = \rho_0 (1 - \beta\Delta T)$, and the specific heat of the incompressible fluid is denoted by $c$. The term, $\dot{q}$ in the last equation denotes the rate of internal heat production per unit volume. At steady-state, the time derivatives vanish and the convective motion is aligned along the direction of the vertical heat flux as the thermal gradient along film thickness is maintained constant. Hence, the spatio-temporal temperature dependence equation from the above reduces to a heat-diffusion equation with a source term. Based on the boundary conditions, the in-plane two-dimensional solutions are harmonic in nature, which explains the appearance of alternating regions of hot and cold domains. As discussed above, for a symmetric rigid-rigid boundary condition, $Ra_c = 1708$. In our experimental methodology, an asymmetric boundary condition (rigid-free) is maintained. Numerically, the $Ra_c$ in such a setting has been computed to be approximately $1100$~\cite{glomski2012precise}.
\section{Methodology}
An IR camera captures the thermal images of the top layer of the fluid-film as the Rayleigh-B{\'e}nard system is driven from a room temperature equilibrium state, maintained at $23^\circ C$, to an out-of-equilibrium steady-state and back. The fluid is placed in the copper pan and it is then heated by regulating the externally applied voltage. Once the power is switched on it takes $2$~hours for the system to reach a steady-state. After the system reaches a steady-state, the external driving is switched off and the system is allowed to cool for another $2$~hours such that the system relaxes back to room temperature. A high viscosity silicone oil is used for the purpose of the current study~\cite{chatterjee2019coexisting,shinetsu}. It is always ensured that a large pan diameter ($2R$) to fluid-film thickness ($l_z$) is maintained, $2R/l_z\simeq~225~mm/5~mm\sim45$, as the goal is to have convection cells over as wide as an area possible for the thermal imaging to yield significant temperature statistics. The thermal dataset thus obtained consists of high-resolution static images and movies that capture the dynamics of the convection cells as they emerge when the system is being driven, and as they dissipate when the system starts relaxing back to room temperature. These images are recorded in grey-scale where each pixel indicates an intensity value that can be later transformed into corresponding temperature values, $T_k$ through a linear interpolation. For lower magnitudes of the external driving field, the convection cells concentrate at the center of the copper pan. This results in an annular region that is devoid of patterns. 

\begin{figure}[t]
    \centering
    \includegraphics[scale=0.8]{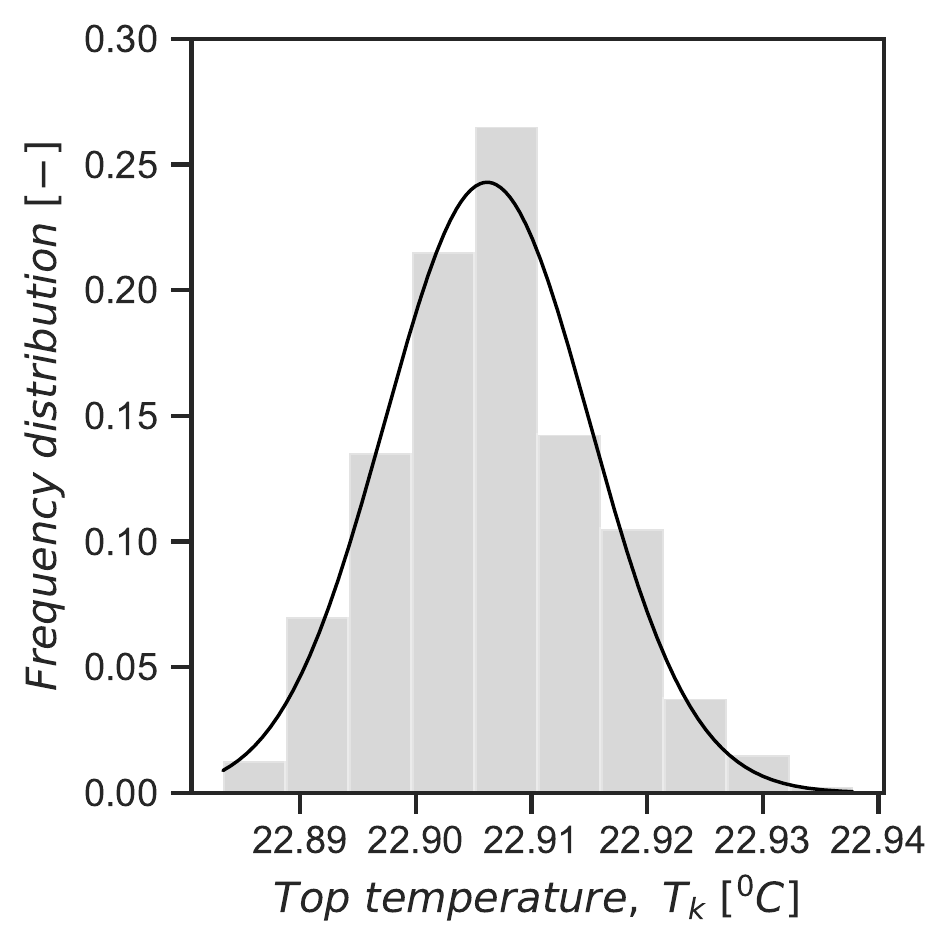}
    \caption{Histogram plot with a Gaussian fit for the room temperature equilibrium state averaged over nine independent trials. The data is obtained after the Rayleigh-B{\'e}nard system relaxes back to room temperature equilibrium state.}
    \label{fig2}
\end{figure}

For the purpose of analysis, regions of interest (ROI) are selected over these images and statistical results are computed. In our previous work, the central region with dense accumulation of convective cells has been denoted by $P$, and the outer annular region by $R$, see for example Figure~5(b) in~\cite{yadati2019spatio}. Further, collection of `hot' and `cold' regions in the steady-state images are obtained by thresholding each image. The image threshold is obtained by spatially averaging the temperature in the annular region at steady-state. The spatially averaged mean temperature over the region of interest is denoted by $\langle T\rangle = \Sigma_k T_k/N$ and the deviation in temperature at each pixel coordinate from the mean is denoted by, $\delta T_k = T_k - \langle T\rangle$. Further, this deviation is scaled by the mean temperature of the region of interest and is denoted by, $t_k = (T_k - \langle T\rangle)/\langle T\rangle$. The statistics for the reference equilibrium state is obtained by analyzing the images which were recorded after the system relaxed back to room temperature equilibrium state. In Figure~\ref{fig2}, we present the data for the reference state from nine independent trial runs after the system relaxes back to room temperature equilibrium. The histogram data is averaged over nine independent trials and fit with a Gaussian distribution function. The mean obtained from the Gaussian fit, $\langle T\rangle = 22.905^\circ C$ is in agreement with the recorded room temperature ($23^\circ C$). Since the camera is essentially recording equilibrium fluctuations (random noise) the standard deviation from the fit is expected to be very low, $\sigma\sim 0.009^\circ C$ which is of the order of magnitude as the sensitivity of the camera. We consider this data as a reference and compare the distribution profiles with those that are obtained from the non-equilibrium steady-state due to external driving. 

The time-averaged data is obtained by observing the dynamics of the Rayleigh-B{\'e}nard convection after a steady-state has been reached. The data is collected by recording a movie for $15$ minutes at $30$ frames per second totaling $27,000$ frames, and statistical analysis is performed on this image dataset by selecting regions of interest as described earlier. In the following section, we discuss the spatial and temporal steady-state statistics in detail. 
\section{Results}
\subsection{Steady-state temporal analysis}
\begin{figure}[t]
    \centering
    \includegraphics[scale=0.45]{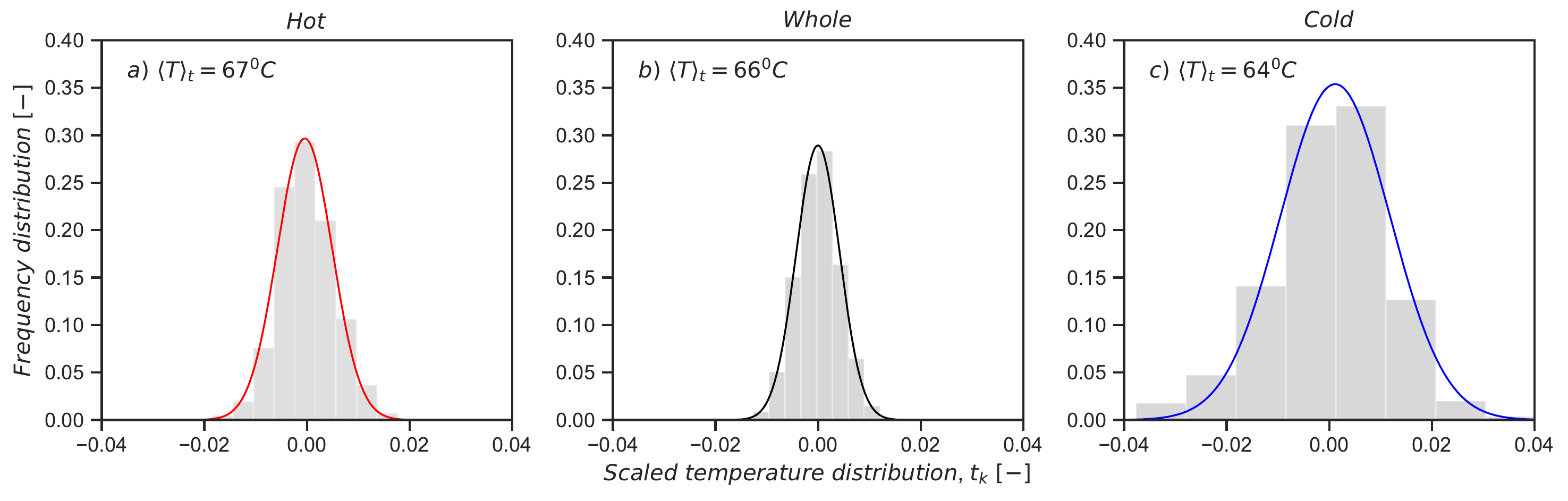}
    \caption{Frequency distribution plots showing scaled thermal data of the top layer at steady-state ($Ra=1410$), temporally averaged over $27,000$ frames or $15$ minutes. In $(a)$ the region of interest is chosen over the hot spots (standard deviation = $5.35\times 10^{-3}$), $(b)$ over the entire top layer of the fluid-film (standard deviation = $4.25\times 10^{-3}$), and $(c)$ over the cold spots (standard deviation = $1.05\times 10^{-2}$) with respective Gaussian fits. The temporally averaged mean temperatures, $\langle T\rangle_t$ of the respective regions of interest are denoted in each panel~\cite{chatterjee2019coexisting}.}
    \label{fig3}
\end{figure}
In Figure~\ref{fig3}, we plot the time-averaged scaled thermal distribution of the top layer of the Rayleigh-B{\'e}nard system after it has reached a steady-state. Figure~\ref{fig3}a shows the frequency distribution histogram with a normal distribution function (in red) centered around the origin for the collection of all the hot spots inside our region of interest. Similarly, Figures~\ref{fig3}b and~\ref{fig3}c denote the frequency distribution histograms and normal distribution curves (also centered around the origin) for the entire region of interest (in black), and all the cold spots (in blue) inside our region of interest, respectively. 

The normality in the distribution plots is not unusual. The thermal data at steady-state behaves as an i.i.d (independent and identically distributed) random variable with a finite mean. Therefore, the time-averaged distribution at a non-equilibrium steady state bears resemblance with that of the temperature distribution profile at room temperature. While all of these states can be described by smooth Gaussian curves and their occurrence can be explained by the Central Limit Theorem, yet there does exist a key difference~\cite{aleksandr1949mathematical}. The standard deviation of the fit shown in Figure~\ref{fig2} differs at least by a scale of magnitude when compared to the fits in Figure~\ref{fig3} (note that the fits shown in Figure~\ref{fig3} are scaled by the mean temperatures). While the profiles of the distribution curves in both cases are qualitatively similar, a rather simple observation - absence of heat flux in the former - can explain this observation.

However, the key takeaway from this result is that the thermal measurements sampled at steady-state (for individual domains and the whole region) bear statistical resemblance to the thermal measurements at equilibrium. Since, the thermal profile at steady-state converges to a Gaussian distribution function the thermal data sampled at every $15$ seconds for the whole region, and isolated domains represent independent random variable similar to i.i.ds for the room-temperature equilibrium case. This implies long-time stability of the convection cells.
\subsection{Steady-state spatial analysis}
\begin{figure}[t]
    \centering
    \includegraphics[scale=0.65]{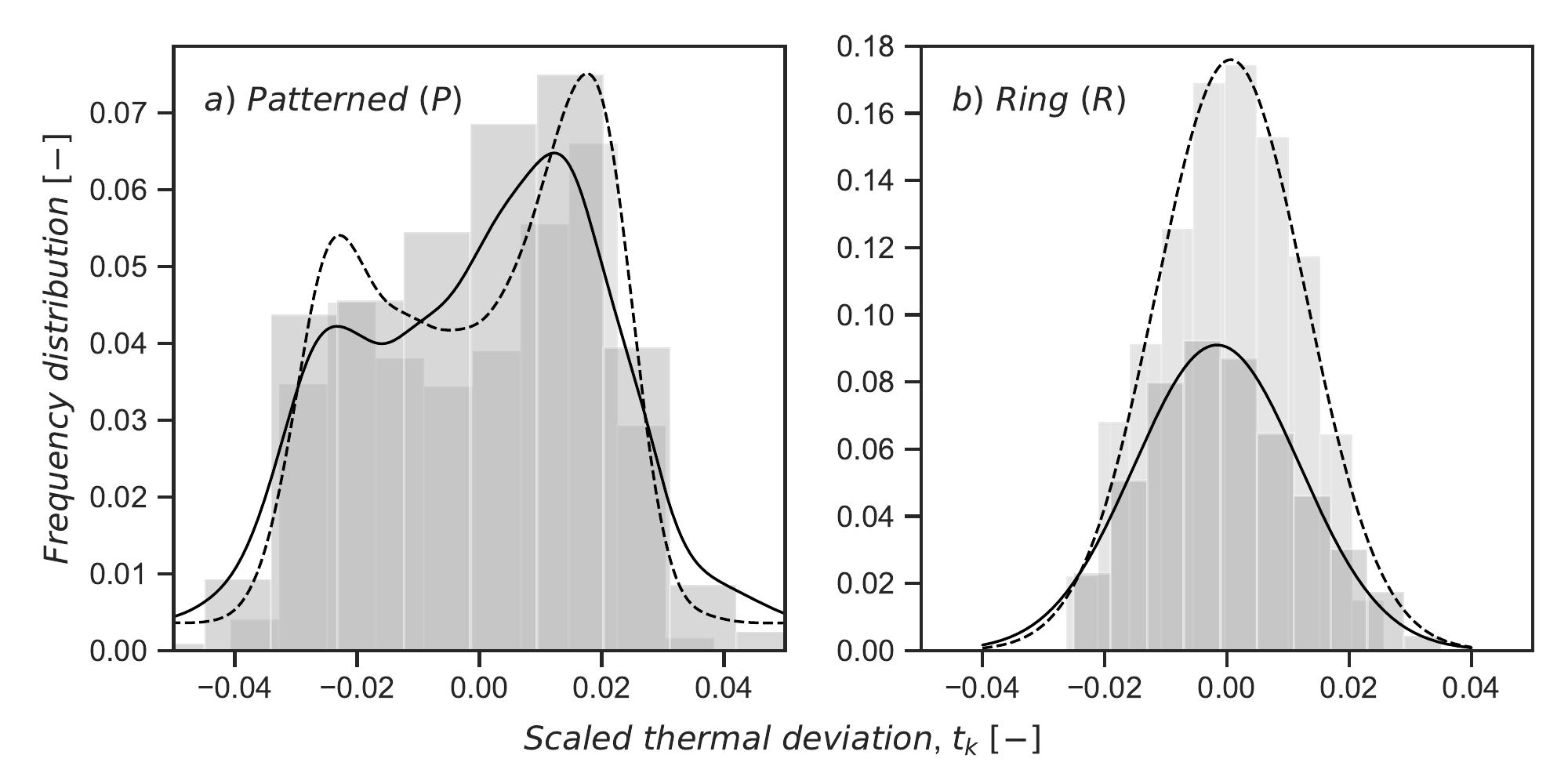}
    \caption{a) Figure shows kernel density estimate plots for the scaled thermal data for the patterned region ($P$) for $Ra = 1410$ (dashed) and $Ra=1670$ (solid), and b) Gaussian fits for the scaled thermal data for the annular region devoid of patterns ($R$). The ROI for the above statistics spans approximately $2500$ pixels.}
    \label{fig4}
\end{figure}
In Figure~\ref{fig4} we plot the spatial distribution of the scaled thermal data for the steady-state images for two samples: $Ra = 1410$ and $1670$. The two regions, $R$ and $P$ are selected as described above. In Figure~\ref{fig4}a the region of interest is the patterned region $P$, and it is visibly evident that the spatial symmetry is broken due to the emergence of patterns/order. The histogram plots and kernel density estimates for the scaled thermal data are shown. The two peaks represent the spatially averaged temperatures of the hot and cold regions. We have also pointed out in our previous work that the asymmetric kernel density estimates shown in Figure~\ref{fig4}a can be characterized by two independent Gaussian fits, which signifies the coexistence of collections of locally equilibrated stable thermal domains in space~\cite{chatterjee2019coexisting,yadati2019spatio}. In the case of the annular region, $R$ the scaled thermal deviation can be characterized by normal distribution functions centered around the origin as shown in Figure~\ref{fig4}b (standard deviation = $1.25\times 10^{-3}$ for $Ra = 1410$, and $1.35\times 10^{-3}$ for $Ra = 1670$). 

The emergence of two peaks at steady-state opens up the avenue to investigate the role of emergent fluxes and the resultant thermodynamic forces. It is near equilibrium, that one can use the linear non-equilibrium thermodynamics postulates. Therefore, it is imperative to consider the functional relationship between them as patterns emerge due to external driving.
\subsection{Force-flux relation and entropy production}
\begin{figure}[t]
    \centering
    \includegraphics[scale=0.65]{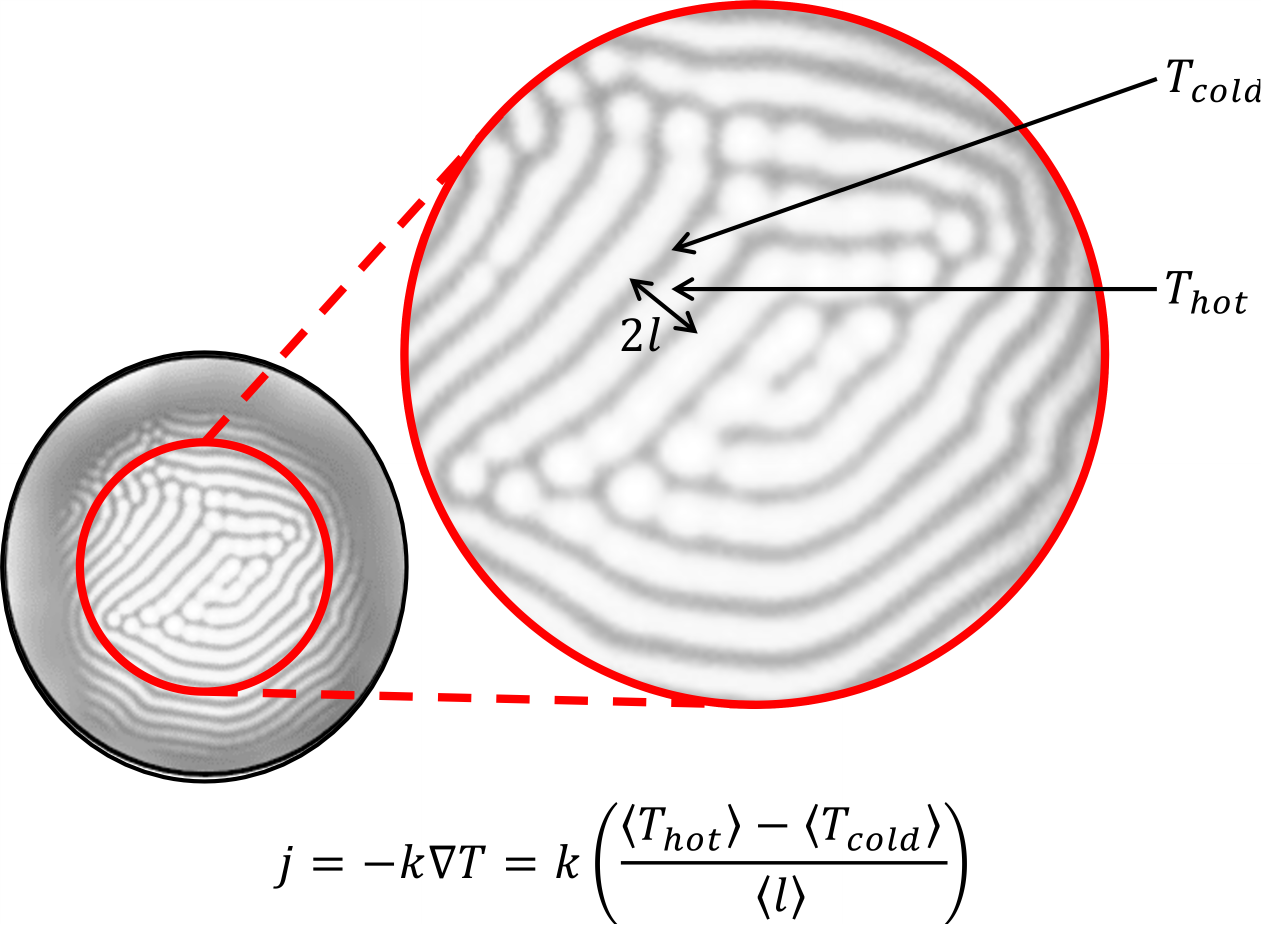}
    \caption{An illustration of the top film IR image of the Rayleigh-B{\'e}nard system at steady-state. The bright regions are `hot' while the darker regions are `cold'. The separation between them is denoted by $l$. The emergent heat-flux is then given by, $j = -k\nabla T = k(\langle T_{hot}\rangle - \langle T_{cold}\rangle)/ \langle l\rangle$ where $k$ is the thermal conductivity of the working fluid. The averaging is performed spatially and the heat-flux is calculated for a sequence of images as the system is driven from room temperature to a non-equilibrium steady-state.}
    \label{fig}
\end{figure}

\begin{figure}[hb!]
    \centering
    \includegraphics[scale=0.7]{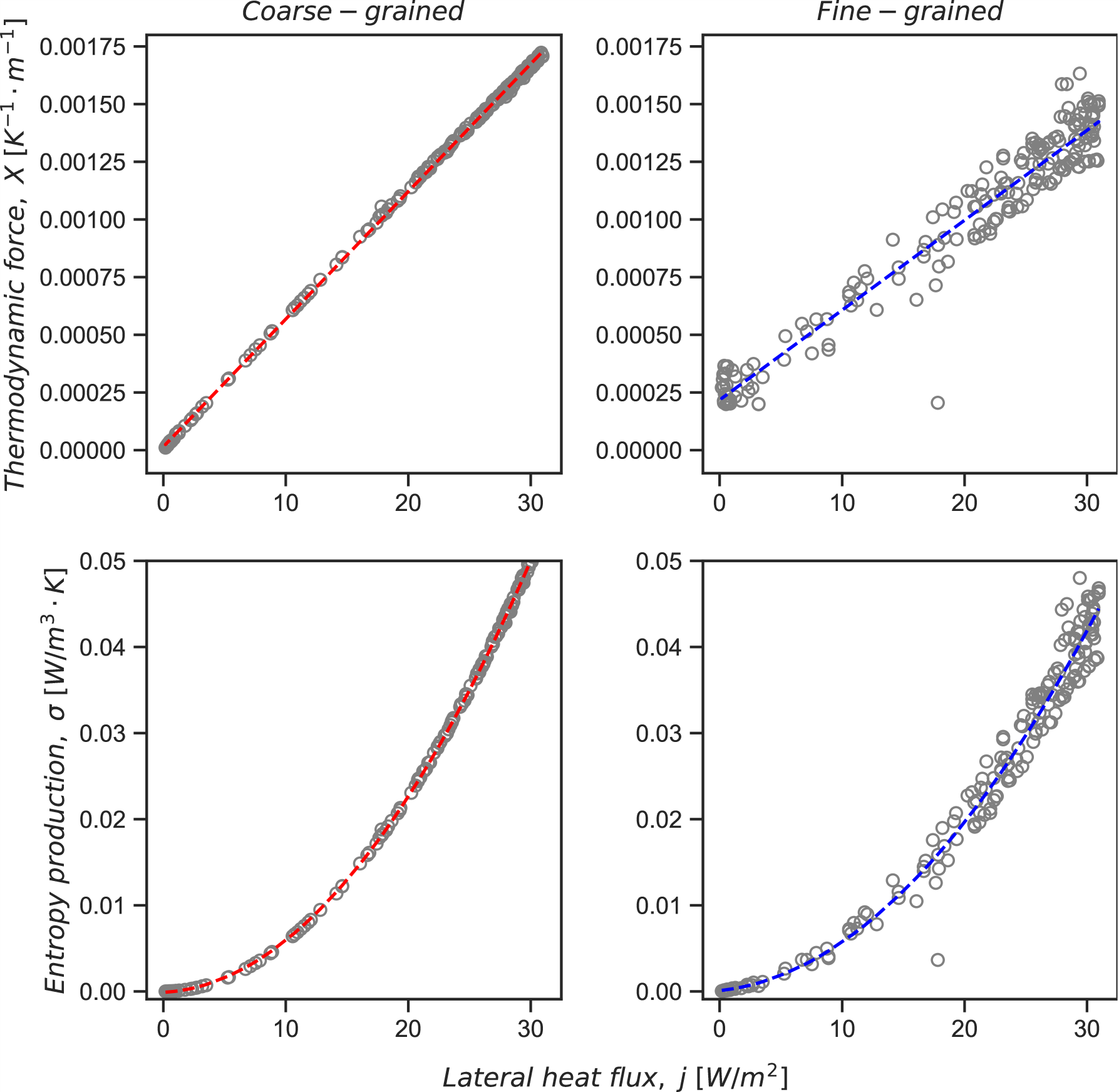}
    \caption{Top panel plots show a linear relationship between thermodynamic force, $X=\nabla(1/T)$ and emergent heat-flux, $j$ in the transition regime. A linear relationship between thermodynamic force and heat-flux implies a quadratic in entropy production given by, $\sigma = jX$, shown in bottom panel. The pair of plots in the left (in red) is computed by coarse-graining the data while the plots on the right (in blue) is computed by performing a fine-grained analysis of the image data.}
    \label{fig5}
\end{figure}
Thermodynamic forces can be understood as the driving forces of transport and reaction phenomena. The steady-state spatial analysis posits that stable domains of high temperature and low temperature emerge as the Rayleigh-B{\'e}nard system is driven out of room-temperature equilibrium state. The gradient in the thermal manifold thus imply the emergence of a lateral heat-flux orthogonal to the direction of the convective motion. This emergent heat-flux, $j$ is computed by dividing the emergent heat differential by the mean separation between hot and cold domains, as shown in Figure~\ref{fig}. The emergent flux, the thermodynamic force, and the entropy production can be expressed by the following set of equations,

\begin{equation}
\begin{gathered}
     j =-k\nabla T =k\left(\frac{\langle T_{hot}\rangle - \langle T_{cold}\rangle}{\langle l\rangle}\right)\\
     X=\nabla(1/T)\\
     \sigma = jX
\end{gathered}
    \label{eqn3}
\end{equation}

The associated thermodynamic force, $X$ is computed in two ways in order to justify the linearity in the flux-force relationship. In the first instance, the thermodynamic force is obtained by coarse graining the image data. Thus, the thermodynamic force can be expressed as, $X=\nabla(1/T) = -\nabla(T)/T^2$. As the emergent heat-flux is written as, $j =-k\nabla T$, the thermodynamic force therefore can be further expressed as, $X=-\nabla(T)/T^2=j/kT^2$. Since, the flux is calculated on the coarse-grained image and is equal to $k\left(\langle T_{hot}\rangle - \langle T_{cold}\rangle\right)/\langle l\rangle$ the thermodynamic force can thus be written as, $\left(\langle T_{hot}\rangle - \langle T_{cold}\rangle\right)/(\langle l\rangle\times T^2)$. The $T$ in the denominator is taken to be the mean temperature of the top film, $\langle T_{top}\rangle$ calculated for each frame at every time-step. As the system is driven from a room temperature equilibrium state to an out-of-equilibrium steady-state, the mean top film temperature is time-dependent in the transition regime. Therefore, the linearity in the flux-force relationship is not trivial in the coarse-grained analysis.

The linear relationship between flux and force is shown in Figure~\ref{fig5}a for the coarse-grained scenario. The slope of the linear fit, $1/k\bar{T}^2$ after substituting appropriate numerical value of the heat transport coefficient, $k$ results into, $\bar{T}\sim 66^\circ C$ which is the mean top-film temperature at steady-state for the sample (see, Figure~\ref{fig3}b for reference). Combining the expressions for force and flux from Equation~\ref{eqn3} we obtain the rate of entropy production per unit volume, $\sigma = jX$. In Figure~\ref{fig5}c, we plot $\sigma$ as a function of the lateral heat-flux for the coarse-grained scenario. As expected, the entropy production follows a quadratic trend.

\begin{figure}[t]
    \centering
    \includegraphics[scale=0.65]{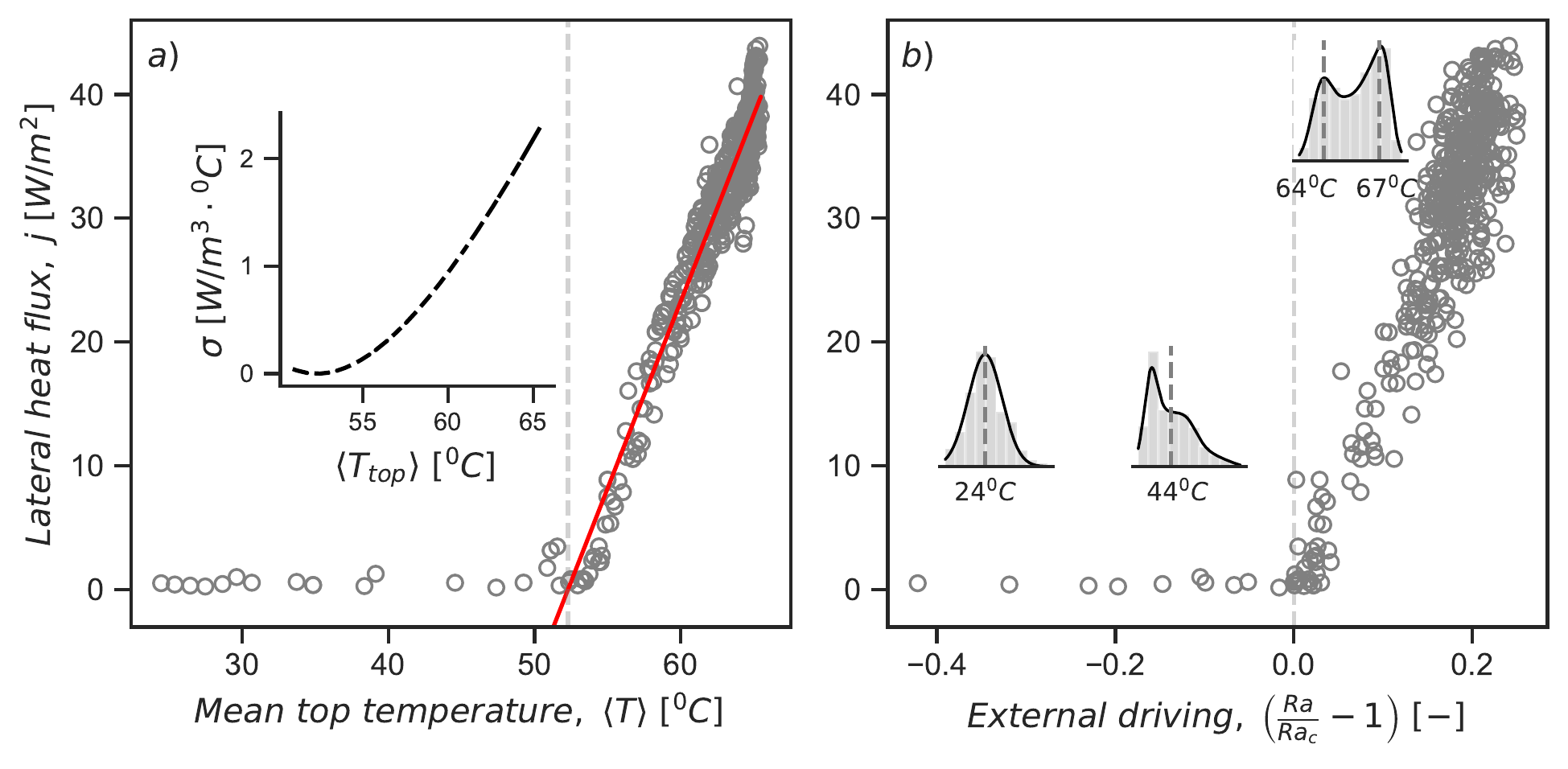}
    \caption{a) Lateral heat flux is plot as a function of the top film temperature. The red solid line depicts a linear fit of the form: $j = a(\langle T\rangle - \langle T_c\rangle)$, where $\langle T_c\rangle\sim 52.2^\circ C$ (shown as a vertical dotted line). Inset plot shows entropy production (per unit volume) as a function of mean top temperature, and b) Lateral heat flux is plot as a function of thermal driving, $(Ra/Ra_c - 1)$ where $Ra_c$ denotes the critical Rayleigh number which corresponds to $\langle T_c\rangle$ as the mean top temperature, signalling the onset of convection. The spatial thermal distribution plots in three different regimes are also shown - room-temperature equilibrium, transient regime, and steady-state.}
    \label{fig6}
\end{figure}

For a fine-grained estimation of the thermodynamic force, we calculate pixel-by-pixel gradient over the image ROI. This is done by choosing an appropriate averaging window, $\alpha$ and then calculating the directional derivatives, $\partial_yT = (T(i, j+\alpha) - T(i, j-\alpha))/(j + \alpha - (j - \alpha))$ and $\partial_xT = (T(i + \alpha, j) - T(i - \alpha, j))/(i + \alpha - (i - \alpha))$. The average of the magnitude of these derivatives over the ROI is then equal to $\nabla T$ for that image. As earlier, this value is then divided by square of the mean top-film temperature, ${\langle T_{top}\rangle}^2$ to obtain the magnitude of the thermodynamic force. In Figure~\ref{fig5}b, the above estimate of the thermodynamic force is plot against the emergent heat-flux which is obtained as described earlier. We can see that there exists a linear relationship between the two. The slope of the linear fit, $1/k\bar{T}^2$ in this case along with the heat transport coefficient is then used to estimate, $\bar{T}\sim 63^\circ C$ which is close to the mean top-film temperature at steady-state for the sample. As earlier, the entropy production is then computed and as expected it follows a quadratic trend as seen in Figure~\ref{fig5}d. 

The lateral heat-flux and the thermodynamic force emerge as patterns appear or convective instability sets in. Therefore, below critical temperature or critical Rayleigh number one should expect no lateral heat flux. In Figure~\ref{fig6}a, we plot the lateral heat-flux as a function of mean top temperature. We can clearly see that the emergent heat-flux is zero upto $\langle T\rangle\sim 52^\circ C$, and following which there is a steep increase till the system reaches a steady-state. The observed trend is fit with a linear function of the form, $j = a(\langle T\rangle - \langle T_c\rangle)$, such that $j=0~\forall~\langle T\rangle\leq\langle T_c\rangle$. The critical top-film temperature obtained from the optimal fit parameters is $\langle T_c\rangle\sim 52.3^\circ C$, shown in the figure as a vertical dotted line. In Figure~\ref{fig6}b, the lateral heat-flux is plot as a function of thermal driving, $\epsilon = \left(Ra/Ra_c - 1\right)$ where the Rayleigh number is computed at every instant in time. The observed relationship between the Rayleigh number and the emergent heat-flux thus obtained from the locally equilibrated domains is very similar to the relationship between the convective heat-flux and the reduced Rayleigh number, $\epsilon$ as previously shown by Meyer et. al.~\cite{meyer1988pattern}. Upon substituting $\langle T_c\rangle$ from the linear fit we obtain $Ra_c\sim 1128$ which is very close to the numerically computed critical Rayleigh number for an asymmetric boundary condition. While we acknowledge that the flux and temperature measurements are not independent, the linearity derived in Figures~\ref{fig5} and~\ref{fig6} cannot be considered strong results yet. Based on previously discussed studies, this work can therefore serve as a starting point for rigorous experimental and theoretical studies in this direction.
\section{Discussion}
In this section, we discuss the conceptual and theoretical implications of our results. But first, we want to emphasize why this result is important. The local equilibrium condition until now has either been assumed as an \textit{a priori} condition during the treatment of non-equilibrium steady-states, or it has been discussed in the literature entirely from either a pure theoretical or a phenomenological approach~\cite{vilar2001thermodynamics,goto2016local,lebon1980extension,serdyukov2018macroscopic}. Moreover, some have even questioned the validity of the local equilibrium assumption~\cite{ben2018validity,ben2020entropy}. This work establishes, for the first time, that the local equilibrium hypothesis is indeed valid and can be experimentally observed in a out-of-equilibrium thermodynamic system. 

We start by examining the conditions for the validity of the local equilibrium hypothesis. Based on extensive previous works, both numerical and analytical it has been suggested that a linear constitutive relationship between thermodynamic fluxes and thermodynamic forces justify the local equilibrium conjecture in non-equilibrium steady-states~\cite{bedeaux2003nonequilibrium,jou1996extended,glavatskiy2009numerical,bedeaux1986nonequilibrium}. Especially in binary mixtures, it has been verified numerically that the interface between the two components at stationary state evaporation and condensation is in local equilibrium~\cite{johannessen2003nonequilibrium}. In the Rayleigh-B{\'e}nard system the two phases - order and disorder emerge as convective instabilities. Therefore, the first criteria to consider is stability of the emergent structures and the nature of the spatio-temporal temperature distribution. Typically, stability and perturbation analysis are carried out numerically for different boundary conditions in order to obtain the critical values of the dimensionless numbers, heat transport, convective velocities etc. From our thermodynamic study, we intend to rather quantify the stability of the emergent phases based on first-order statistics as a preliminary development. Qualitatively, the distribution profiles for the reference equilibrium state in Figures~\ref{fig2} and non-equilibrium steady-states in Figures~\ref{fig3} and~\ref{fig4}b are Gaussian fits. The only quantitative difference being the standard deviation of the reference fit is atleast one order of magnitude less than the steady-state fits. As we discussed earlier this can be explained by the absence of an external heat flux in the former. Moreover, the convergence of the thermal data sampled at steady-state to Gaussian distributions imply that the data is independent and identically distributed in the respective samples. This is in contrast to the transient regime or the spatial ROI consisting of both hot and cold domains wherein the temperature data is neither independent nor identically distributed. Especially in the context of the local equilibrium, this allows one to derive phenomenological laws of non-equilibrium thermodynamics in the limit of small thermal fluctuations, and especially in the Gaussian limit where means and modes of the distribution coincide~\cite{lavenda2019nonequilibrium}. This observation ensures long-term temporal stability of the system and of the spatially distributed thermal patterns at steady-state. Indeed, this definition of stability does not conform with the conventional definitions of stability of a dynamical system but our preliminary results do indicate that the temperature of the system at steady-state converges to a stable limit-cycle.

The next criterion is concerned with the time-scales that dictate the steady-state evolution of the system, and the emergence of spatio-temporal thermal patterns. A Rayleigh-B{\'e}nard system in the non-turbulent regime, exhibits atleast two prominent non-overlapping time-scales: a macroscopic timescale, $\tau_M$ responsible for the equilibration of locally stable spatio-temporal thermal patterns, and a microscopic time-scale, $\tau_m$ dictating the steady-state evolution of the system. The two time-scales are obtained by fitting the time evolution of the emergent (orthogonal) heat-flux and the convective (vertical) heat-flux by a generic function of the form: $j(t) = j_0(1-\exp(-(t-t^\prime)/\tau))$, where $\tau$ represents the characteristic time constants (relaxation time) for the two cases, and $t^\prime$ the delay. The relaxation time or the experiment time for the system to achieve a steady-state is much faster than the observed time responsible for the relaxation of the thermal patterns. Thus the ratio, $De=\tau_m/\tau_M = (\sim 500~\text{sec}/3600~\text{sec})$ is found to be $\sim 0.1$ for our experimental run. Note, $t^\prime=0$ for the convective heat-flux whereas, $t^\prime\neq 0$ for the emergent heat-flux as structures emerge once $Ra_c$ is achieved. As, noted earlier a $De<<1$ justifies the use of local equilibrium postulate as the two time-scales are well separated. Finally, a necessary condition for the local equilibrium hypothesis to hold true is the linearity in the force-flux relation. In Figure~\ref{fig5}, we show the linear relationship between the resultant thermodynamic force and the emergent heat-flux. We also calculate the rate of entropy production as stable spatio-temporal thermal patterns emerge. The quadratic nature of the curves in Figures~\ref{fig5}c and~\ref{fig5}d indicate that the system drifts to a state of higher entropy. Moreover, the Rayleigh-B{\'e}nard system gives us the control in tuning the system's response to external driving. This leads us to an important question: at what point does the local equilibrium hypothesis fail? Qualitatively, larger magnitudes of $\epsilon$ would drive the system to a state of instability wherein the linearity in the force-flux relation would be eventually broken. Currently, we lack sufficient experimental evidence to support this conjecture. 

The validity of the local equilibrium hypothesis has an important consequence. As a result of which thermodynamic state variables (in this case, temperature) can be defined and principles from equilibrium thermodynamics can be used to describe these states quantitatively~\cite{muschik2020discrete}. The idea is conceptually similar to a discrete system where the out-of-equilibrium system interacts with the environment at equilibrium, such that a measurable `contact temperature' can be defined. Indeed, there are numerous situations in thermal sciences where such situations arise. This does beg a philosophical question: if, sufficiently close to equilibrium, one can \textit{only} measure the standard thermodynamic state variables $P,V,T...$ then how can one tell whether the system is actually at equilibrium or it has been driven out-of-equilibrium to a meta-stable state with emergent complexity? As a~\textit{gedankenexperiment} consider an inertial frame of reference as an analogue for the room temperature equilibrium state. An inertial frame of reference describes time and space homogeneously, isotropically, and in a time-independent manner, identical to the characteristics that define an equilibrium thermodynamic state. Similarly, a non-inertial frame of reference that accelerates relative to an inertial frame finds an analogue in a thermodynamic state of a system that is being driven out-of-equilibrium. While a frame of reference in relativistic mechanics is described by the space-time coordinates, a thermodynamic state is represented by a set of intensive variables in the phase-space. A thermodynamic state that is being driven out-of-equilibrium then can be characterized not only by the set of thermodynamic state-variables but also by their spatio-temporal derivatives (or gradients). For a system at a non-equilibrium steady-state the time derivatives vanish, while the spatial derivatives (gradients) exist. The measured mean temperature of the system will indicate whether the system is at room temperature or has been driven out-of-equilibrium, and the width of the distribution will denote the presence (or absence) of an external energy flux. But none of the statistics would indicate the spontaneous emergence of spatio-temporal patterns. The spatial symmetry (isotropy) is broken \textit{only} when locally equilibrated domains emerge. Thus, for an uninformed observer (akin to the IR camera - an informed observer) it is impossible to make a distinction between an equilibrium state at room temperature and a non-equilibrium meta-stable state that consists of spatially distributed stable patterns solely on the basis of measurements made in local regions in space. Since these two `frames of reference' are governed by the same principle of statistical mechanics, for an uninformed observer they must be thermodynamically identical. However, when the observer can `see' the thermal landscape, they can clearly distinguish the two states (and perceive the notion of emergent forces and fluxes) - a viewpoint which is analogous to Einstein's weak principle of equivalence in relativity~\cite{einstein1905tragheit,einstein1935particle}. 

Thinking about a thermodynamic system from the point of view of a landscape (or a manifold) brings us closer to develop a theoretical approach in quantifying emergent order in non-equilibrium steady-states where local equilibrium is satisfied~\cite{chatterjee2019many,chatterjee2019coexisting,van2017guyer}. Being able to define a steady-state non-equilibrium state variable precisely, allows us the freedom to quantify entropy production due to heat fluxes and calculate thermodynamic forces~\cite{onsager1931reciprocal,lucia2020time,lucia2020time1,chatterjee2020time,riek2020entropy}. The validity of the current work also sets the stage to apply variational methods to describe the evolution of non-equilibrium steady-states, study bistability, bifurcation, and test the MEPP in Rayleigh-B{\'e}nard and other prototypical irreversible systems~\cite{martyushev2006maximum,de2019oscillatory,ban2019thermodynamic,ban2020thermodynamic}. This further opens up the avenue for the first time to define and measure extensive variables like, work, heat, and their inter-conversion in a non-equilibrium steady-state - a precursor to a first principles' generalized equation of state that takes into account emergent order~\cite{chatterjee2020studies}.
\section{Concluding Remarks}
In this paper, we study a classical irreversible system and examine the validity of the local equilibrium hypothesis. We discuss the role of thermodynamic forces as fluxes emerge with the onset of thermal convection. Based on the linear flux-force relation, well separated time-scales, and thermal statistics obtained from long-term spatio-temporal stability of the thermal patterns, we validate the local equilibrium conjecture. We discuss in detail the importance of this result in the study of non-equilibrium thermodynamics, and consider possible future directions.
\section{Acknowledgments}
The authors are extremely grateful to the reviewers and the editor for their invaluable feedback and constructive criticism which have greatly improved the paper.






\end{document}